\documentclass[utf8]{FrontiersinHarvard}% 

\usepackage{url,hyperref,microtype,subcaption}
\usepackage[onehalfspacing]{setspace}
\usepackage{booktabs}

\def\keyFont{\fontsize{8}{11}\helveticabold }
\def\firstAuthorLast{Jones } %use et al only if is more than 1 author
\def\Authors{K. L. Jones\,$^{1,*}$, J. Kovoor\,$^{1}$ R. Kanungo\ $^{2,3}$}
\def\Address{$^{1}$Department of Physics and Astronomy, University of Tennessee, Knoxville, TN, 37996, USA\\
 $^{2}$ Astronomy and Physics Department, Saint Mary's University, Halifax, Nova Scotia, B3H 3C3, Canada\\
$^{3}$TRIUMF, Vancouver, British Columbia, V6T 4A3, Canada }
\def\corrAuthor{K. L. Jones}

\def\corrEmail{kgrzywac@utk.edu}

\begin{document}
\onecolumn
\firstpage{1}

\title {Status of experimental knowledge on the unbound nucleus $^{13}$Be} 

\author[\firstAuthorLast ]{\Authors} %This field will be automatically populated
\address{} %This field will be automatically populated
\correspondance{} %This field will be automatically populated
\extraAuth{}% If there are more than 1 corresponding author, comment this line and uncomment the next one.
%\extraAuth{corresponding Author2 \\ Laboratory X2, Institute X2, Department X2, Organization X2, Street X2, City X2 , State XX2 (only USA, Canada and Australia), Zip Code2, X2 Country X2, email2@uni2.edu}
\maketitle

\begin{abstract}

%%% Leave the Abstract empty if your article does not require one, please see the Summary Table for full details.
\section{}
The structure of the unbound nucleus $^{13}$Be is important to understanding the Borromean, two-neutron halo nucleus $^{14}$Be. The experimental studies conducted over the last four decades are reviewed in the context of the beryllium chain of isotopes and some significant theoretical studies. One focus of this paper is the comparison of new data from a $^{12}$Be(d,p) reaction in inverse kinematics, which was analyzed using GEANT4 simulations and a Bayesian fitting procedure, with previous measurements. Two possible scenarios to explain the strength below 1~MeV above the neutron separation energy were proposed in that study: a single $p$-wave resonance, or a mixture of an $s$-wave virtual state with a weaker either $p$-wave or $d$-wave resonance. Comparisons of recent invariant mass and the (d,p) experiments show good agreement between the transfer measurement and the two most recent high-energy nucleon removal measurements.

\tiny
 \keyFont{ \section{Keywords:} Direct Reactions, Clustering, Beryllium, $^{13}$Be, $^{12}$Be} %All article types: you may provide up to 8 keywords; at least 5 are mandatory.
\end{abstract}

\section{Introduction}

%For Original Research Articles \citep{conference}, Clinical Trial Articles \citep{article}, and Technology Reports \citep{patent}, the introduction should be succinct, with no subheadings \citep{book}. For Case Reports the Introduction should include symptoms at presentation \citep{chapter}, physical exams and lab results \citep{dataset}.
With just four protons, the particle-bound members of the beryllium chain of isotopes stretch from $^7$Be ($N/Z=0.75$) on the proton-rich side of stability, to $^{14}$Be ($N/Z=2.5$), the two-neutron halo on the neutron-rich side. Adding a single neutron to $^7$Be results in the $^8$Be system of two $\alpha$ particles, which is unbound by only 92~keV. Adding a second neutron produces $^9$Be, the only beryllium isotope that is stable against $\beta$ decay. This stability is a product of two phenomena that occur across the beryllium chain of isotopes, molecular structures \cite{Von96, Von97, Sey81}, and core excitation \cite{Par05}. The molecular structure is closely connected to the Borromean nature of $^9$Be; the three-body system of $\alpha-\alpha-n$ is bound despite both of the two-body subsystems, $\alpha-n$, and $\alpha-\alpha$ being unbound. The delocalized neutron in $^9$Be can be viewed as being exchanged between the $\alpha$ particles \cite{Fre18}. The neutron in the ground state of $^9$Be is well understood as being in a $\pi$-type orbital, in analogy with atomic molecules, whereas the first excited state may be better described with a neutron in the $\sigma$-type orbital. These molecular orbits are intimately related to the prolate nature of the two-$\alpha$ cluster structure \cite{Can20}. 

The cluster structure appears to weaken in $^{10}$Be, as evidenced by the reduced size of its charge radius \cite{Nor09, Kri12}. The two-$\alpha$ plus two-neutron structure is more apparent in excited states closer to the $\alpha$ and neutron separation energies, such as the isomeric second $0^+$ state at $E_x=6.179$~MeV, which has a small $\gamma$ branch to the more compact first $2^+$ state \cite{Fre18}. 

Adding another neutron makes the archetypal one-neutron halo nucleus $^{11}$Be.  A 16\% core-excited component in the ground state of $^{11}$Be was required to reproduce the results from the $^{11}$Be(p,d)$^{10}$Be reaction \cite{For99}. This is much less than for $^{9}$Be, where the core excited component was calculated as around half of the ground state wave function \cite{Par05}. Additionally, dynamical core excitation needs to be included in calculations of both transfer \cite{Del16} and break up \cite{Mor12} reactions.

The parity inversion in $^{11}$Be, where the $1/2^+$ ground state dips 320~keV below the only other bound state with $J^{\pi}=1/2^-$, along with the large collectivity in $^{10}$Be, led to questions about the robustness of the $N=8$ shell closure at $^{12}$Be. Using a three-body model with core excitation, Nunes et al were able to show an increased sphericity in the core $^{10}$Be within $^{12}$Be compared that seen in $^{11}$Be \cite{Nun02}. This in turn led to greater mixing between the $p-$ and $sd-$shell valence neutron states and a melting of the $N=8$ shell closure. The coupling of a $d-$wave neutron with the excited 2$^+$ $^{10}$Be core severely restricts the formation of a halo in $^{12}$Be \cite{Nun05}. Notably, $^{12}$Be is not Borromean, as the n-$^{10}$Be system is bound but is still well described by three-body models.

%With a two-neutron separation energy of 3.67~MeV 12Be
The $N=10$ isotope $^{14}$Be presents the heaviest particle-stable beryllium isotope. The naive shell model would predict a $d_{5/2}$ dominated ground state wave function for $^{14}$Be. However, with the level inversion seen in the other neutron-rich beryllium isotopes, some low-lying $s_{1/2}$ strength is expected.  With two neutrons in the halo, and with significant $\ell >0$ components of the wave function, the halo of $^{14}$Be is much more contained than that of $^{11}$Be, despite being closer to the drip line. It would seem natural to study $^{14}$Be in a three-body model, with $^{12}$Be as a core and two valence neutrons \cite{Des95, Tho96}. Thompson and Zhukov found that adding an $s$-wave virtual state below the well know $\frac{5}{2}^+$ state made $^{13}$Be bound, in contradiction to its non-observance in fragmentation reactions. Reducing the energy, i.e. increasing the scattering length, of the virtual state resulted in the binding energy of $^{14}$Be being too low. The three-body approach of Descouvemont found that only $66\%$ of the ground state wave function of $^{14}$Be could be described as $^{12}$Be $+n+n$  \cite{Des95}.  Labiche et al \cite{Lab99}, using the model of Vinh Mau and Pacheco \cite{Vin96} found that assuming a $\frac{1}{2}^-$ ground state for $^{13}$Be, consistent with the melting of the $N=8$ shell closure seen in $^{11}$Be and $^{10}$Li could reproduce the measured properties of $^{14}$Be. 

Beyond the neutron drip line, $^{15}$Be has been observed to decay to $^{12}$Be through unbound states in $^{14}$Be \cite{Spy11}. The last isotope to be observed is $^{16}$Be, which is bound with respect to one neutron emission, but unbound to the emission of two neutrons \cite{Spy12}. The two neutrons from the decay were observed in a small emission angle.

\section{The unbound nucleus $^{13}${B\lowercase{e}}}
Theoretical studies of $^{13}$Be have used the shell model \cite{Pop85} or a potential model \cite{For19}, the Nilsson model \cite{Mac18}, microscopic cluster models \cite{Des94}, antisymmetrized molecular dynamics \cite{Kan12}, and relativistic mean field theory \cite{Ren97}. As shown in Table \ref{theory}, with the exception of the earliest work, the calculations agree on the existence of a $5/2^+$ resonance, between 2~MeV and 2.5~MeV above the neutron threshold. However, the location of the ground state relative to the neutron threshold is disputed, as is the parity of the ground state. There have also been reaction theory studies of, and comparing to, experimental data, for example, \cite{Bon12} and references therein. \cite{Cas20}, used a transfer to the continuum model including deformation in the $^{12}$Be+n potential, following the prescription of Thompson et al. \cite{Tho04}, to interpret the data from Corsi et al. \cite{Cor19}. This work indicates a $p$-wave resonance at between 0.4 and 0.5~MeV above the threshold.

\begin{table}
\caption{\label{theory} The energy (and $J^{\pi}$ assignments) of low-lying states in $^{13}$Be according to a selection of theoretical studies. Strength 2 refers to any virtual states or resonances around 2~MeV above the neutron threshold.}
%\begin{ruledtabular}
\begin{minipage}{\textwidth}
\begin{center}
\begin{tabular}{lcc}
\toprule
\multicolumn{1}{c}{}&\multicolumn{2}{c}{Energy above the neutron threshold (MeV)}\\
\multicolumn{1}{c}{}&\multicolumn{2}{c}{ }\\
Author (Year)  		& Ground state  											& Strength 2  		\\ \midrule
\cite{Pop85}  		& 1.16 $\left(\frac{1}{2}^-\right)$	 \footnote{Poppelier also calculates a $\left(\frac{5}{2}^+\right)$ state at 1.21~MeV }						
																		&2.44$\left(\frac{5}{2}^-\right)$  \\
Lenske \cite{Ost92}	& 0.9 $\left(\frac{1}{2}^+\right)$									&2.3 $\left(\frac{3}{2}^-\right)$ and 2.45$\left(\frac{5}{2}^+\right)$	\\
\cite{Des94} v2 		& -0.009 $\left(\frac{1}{2}^+\right)$								&2.02 $\left(\frac{5}{2}^+\right)$			\\
\cite{Des94} v4 		& -0.038 $\left(\frac{1}{2}^+\right)$								&2.05 $\left(\frac{5}{2}^+\right)$			\\
%\cite{Bon12}		& $\left(-0.8~ {\rm fm}^{-1}\right)$ \footnote{Bonaccorso also calculation a 	$\left(\frac{1}{2}^-\right)$ state at 0.67~MeV }					
%																		&2.00 $\left(\frac{5}{2}^+\right)$			\\
\cite{For19}		& 0.86 $\left(\frac{1}{2}^+\right)$								&2.11$\left(\frac{5}{2}^+\right)$			\\
%\cite{Mac18}	& 0.0 $\left(\frac{1}{2}^+\right)$		&-							&1.8$\left(\frac{5}{2}^+\right)$	\\
\bottomrule
\end{tabular}
%\end{ruledtabular}
\end{center}
\end{minipage}
\end{table}

There have been many experiments on $^{13}$Be since the first discovery of a resonance \cite{Ale83} at 1.8~MeV above the neutron threshold. Some of the experimental results from the last four decades are collated in Table \ref{Exp}. The reactions used to probe the structure of halo nuclei can be broadly divided into missing mass and invariant mass techniques. Transfer reactions where excitation energies are found from the reaction Q-values, fall into the missing mass category. Knockout, breakup, and Coulomb dissociation, where the final state is reconstructed from two or more fragments, represent invariant mass techniques. In the case of $^{13}$Be the fragments are $^{12}$Be and a neutron. The early measurements, e.g. \cite{Ale83, Ost92, Von95, Bel98} mostly populated $^{13}$Be through multinucleon transfer reactions. An exception is the $^{12}$Be(d,p) experiment performed by \cite{Kor95} at RIKEN. At a beam energy of 55~AMeV, the conditions are not well-matched to observe low $\ell$ transfer, and the carbon in the target largely masked any structure below 2~MeV. Indeed, none of the experiments before the fragmentation experiment at the National Superconducting Cyclotron Laboratory \cite{Tho00} revealed any structure below the $\frac{5}{2}^+$ resonance at 2~MeV.

\begin{table*}
\begin{minipage}{\textwidth}
\caption{\label{Exp} Previous studies of the low-lying structure of $^{13}$Be, up to around 2.5 MeV above the neutron threshold. A strength is defined as a single resonance, or a virtual state, or a mixture of resonances, or a virtual state and a resonance. All experiments found some strength below 0.5~MeV above the neutron threshold, and that is labeled as Ground State in this table. Strength 1b is around 0.5 - 0.9~MeV above the neutron threshold. Strength 2 is the well-known $d$-wave resonance at around 2~MeV above the neutron threshold. Where the literature reports scattering length instead of energy this is quoted in parentheses.}
%\begin{ruledtabular}
\begin{tabular}{llccc}
 \toprule
 \multicolumn{1}{l}{Author (year)}&\multicolumn{1}{l}{Reaction}  & \multicolumn{3}{c}{Energy above the threshold (MeV) or (a$_s$) and J$^{\pi}$}\\
  \multicolumn{1}{l}{}&\multicolumn{1}{l}{}				&  \multicolumn{3}{c}{}\\
 	           &									&Ground State	( Strength 1a)									&Strength 1b 				&Strength 2\\ 
\midrule
 \cite{Ale83}&	$^{14}\text{C}+^{7}\text{Li}$ 			&-												&-						&$1.8$ \\
 \cite{Ost92}&	$^{13}\text{C}+^{14}$C&				&-												&$2.01 \left(\frac{5}{2}^+\text{or }\frac{1}{2}^-\right)$\\
 \cite{Kor95}&	$^{12}$Be+d 						&-												&-						&$2.0$\\
 \cite{Von95}&	$^{13}\text{C}+^{14}$C				&-												&-						&$2.01 \left(\frac{5}{2}^+\right)$\\
 \cite{Bel98}&	$^{14}$C+$^{11}$B					&-												&$0.80$					&$2.02$\\
 \cite{Tho00}&	$^{9}$Be$+^{18}$O					&$0.20\left(\frac{1}{2}^+\right)$ 							&$0.80(\frac{1}{2})$			&$2.02\left(\frac{5}{2}^+\right)$ \\
 \cite{Lec04}&	$^{14}$B+C						&-												&$0.7\left(\frac{1}{2}^+\right)$	&$2.4\left(\frac{5}{2}^+\right)$\\
 \cite{Sim07}&	$^{14}$Be+C					 	&$\left(-3.2~{\rm fm}^{-1}\right) \left(\frac{1}{2}^+\right)$		&-						&$2.00\left(\frac{5}{2}^+\right)$ \\
 \cite{Kon10}&	$^{14}$Be+p						& $\left(-3.4~{\rm fm}^{-1}\right) \left(\frac{1}{2}^+\right)$		&$0.51\left(\frac{1}{2}^-\right)$	&$2.39\left(\frac{5}{2}^+\right)$\\
 \cite{Aks13}&	$^{14}$Be+p						&$0.46\left(\frac{1}{2}^+\right)$							&-						&$1.95\left(\frac{5}{2}^+\right)$\\
 \cite{Ran14}&	$^{14,15}$B+$^{\text{nat}}$C			&$0.40\left(\frac{1}{2}^+\right)$							&$0.85\left(\frac{5}{2}^+\right)$	&$2.35\left(\frac{5}{2}^+\right)$\\
 \cite{Mar15}&	$^{13}$B+$^{9}$Be					&-												&$0.73\left(\frac{1}{2}^+\right)$	&$2.56\left(\frac{5}{2}^+\right)$\\
 \cite{Rib18}&	$^{14}$B+CH$_2$					&$0.44\left(\frac{1}{2}^-\right)$							&$0.86\left(\frac{1}{2}^+\right)$	&$2.11\left(\frac{5}{2}^+\right)$\\
 \cite{Cor19}&	$^{14}$Be+p						&$\left(-9.2~{\rm fm}^{-1}\right) \left(\frac{1}{2}^+\right)$		&$0.48\left(\frac{1}{2}^-\right)$	&$2.30\left(\frac{5}{2}^+\right)$\\
 \cite{Kov23} single&	$^{12}$Be+$^{\rm solid}{\rm D}$ 	&$0.55\left(\frac{1}{2}^-\right)$							&-						&$2.22\left(\frac{5}{2}^+\right)$\\
\multicolumn{1}{l} {\cite{Kov23} mix}&	\multicolumn{1}{l}{$^{12}$Be+$^{\rm solid}{\rm D}$}	
	&\multicolumn{2}{c}{$0-1~{\rm MeV}\left(\frac{1}{2}^+\right) $and$ \left(\frac{1}{2}^-\right)$ \footnote{order undetermined, see text.}}		&\multicolumn{1}{c}{$2.22\left(\frac{5}{2}^+\right)$}\\
 \bottomrule
\end{tabular}
%\end{ruledtabular}
\end{minipage}
\end{table*}

Most of the more recent experiments used invariant mass techniques at energies between 40 AMeV and 400 AMeV. Of these, all except one employed nucleon removal, either a neutron from $^{14}$Be \cite{Sim07, Kon10, Aks13, Cor19} or a proton from $^{14}$B \cite{Lec04, Ran14,Rib18}.  \cite{Mar15} used a nucleon exchange reaction to populate unbound states in $^{13}$Be from a $^{13}$B beam. Spectra from invariant mass methods contain information relating to the initial state of the target (the beam in an inverse kinematics reaction) and reflect final state interactions between outgoing fragments. There can also be complications relating to the reaction mechanism, which can be diffractive or absorptive. 

\section{Recent transfer reaction measurement}
The most recent experiment, \cite{Kov22, Kov23}, used a $^{12}$Be beam from ISAC-II, and the solid deuterium target and detector system IRIS \cite{Kan14} to perform a one-neutron transfer reaction.  The advantages of using the (d,p) reaction at low energy ($E_{beam}=9.5$~AMeV) are that the energy and angular momentum matching conditions are ideal for populating low-$\ell$, near-threshold states. The non-Gaussian experimental response caused by increasingly poor resolution for lower energy protons exiting the solid deuterium target, necessitated an analysis technique that included both simulation and Bayesian fitting methods, as demonstrated in \cite{Hoo21}. The initial analysis included the $\frac{5}{2}^+$ resonance at around 2~MeV and a single either $s$-, $p$-, or $d$-wave virtual state or resonance closer to the threshold, referenced as `single' in Table \ref{Exp}. Additionally, mixes of two out of $s$-, $p$-, and $d$-wave strengths were included to allow for two resonances, or a resonance and a virtual state below the 2~MeV $d$-wave resonance, referenced as `mixed' in Table \ref{Exp}. The fit with the lowest $\chi^2/\rm{NDF}$ for the region below 1~MeV in resonance energy, for both the single and the mixed case, are the ones quoted in Table \ref{Exp}. The rest of this paper relates to the findings in \cite{Kov23} including the comparisons with recent invariant mass measurements \cite{Kon10, Ran14, Rib18, Cor19}. 

\begin{figure}[h!]
\begin{center}
\includegraphics[width=15cm]{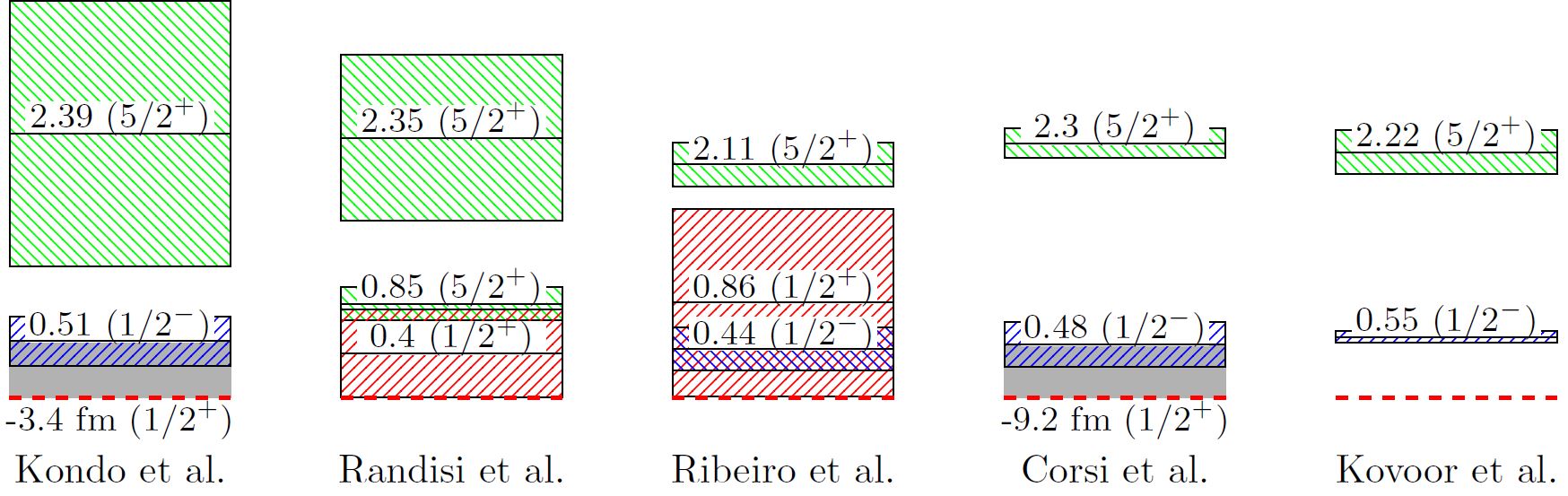}
\end{center}
\caption{Comparison of low-lying structure in $^{13}$Be according to recent invariant mass measurements \cite{Kon10, Ran14, Rib18, Cor19} and a recent transfer reaction measurement \cite{Kov23}. Only the case with a pure $p$-wave is shown for \cite{Kov23} as the position of the waves in the mixtures could not be resolved. Red, blue, and green lines depict $s$-, $p$-, and $d$- waves respectively. The red-dashed lines show the threshold. The gray-shaded region shows the presence of a virtual state. }\label{fig:1}
\end{figure}

A comparison of the transfer measurement (for the case assuming a single state below 2~MeV) with some recent invariant mass measurement results is shown in Figure \ref{fig:1}.  Discrepancies clearly persist in the low-lying structure of $^{13}$Be, even between recent measurements.  The $\frac{5}{2}^+$ state is in a similar position in all the measurements, however, the width is much larger according to the works of \cite{Kon10} and \cite{Ran14}. Similarly, \cite{Ran14} and \cite{Rib18} show a broad $\frac{1}{2}^+$ structure below 1~MeV, whereas \cite{Kon10} and \cite{Cor19} et al. agree on an s-wave virtual state. There is some consistency in the presence of a p-wave resonance around 0.5~MeV, with the exception of \cite{Ran14}. This resonance is noticeably narrower ($\Gamma=0.11$~MeV) in the work of \cite{Kov23}.

The mixed case in \cite{Kov23} is not shown in Figure \ref{fig:1} as the ordering of the $s$- and $p$-wave strength cannot be extracted from the data, only the relative intensities. The  lowest $\chi^2/\rm{NDF}$ was found for a mixture of $s-\left(70^{+8}_{-6}\%\right)$  and $p-\left(30^{+6}_{-8}\%\right)$ wave strengths. It should be noticed that the $\chi^2$/\rm{NDF} (3.87) for the $s-\left(93^{+2}_{-2}\%\right)$  and $d- \left(7^{+2}_{-2}\%\right)$wave mixture below 1~MeV is similar to that for the $s$- and $p$-wave mixture (3.32). Therefore, the initial conclusions of the study were that the near-threshold strength was either pure $p$-wave or a mixture, dominated by an $s$-wave strength, with a weaker resonance of either a $p$- or $d$-wave nature.

\begin{figure}[h!]
\begin{center}
\includegraphics[width=14cm]{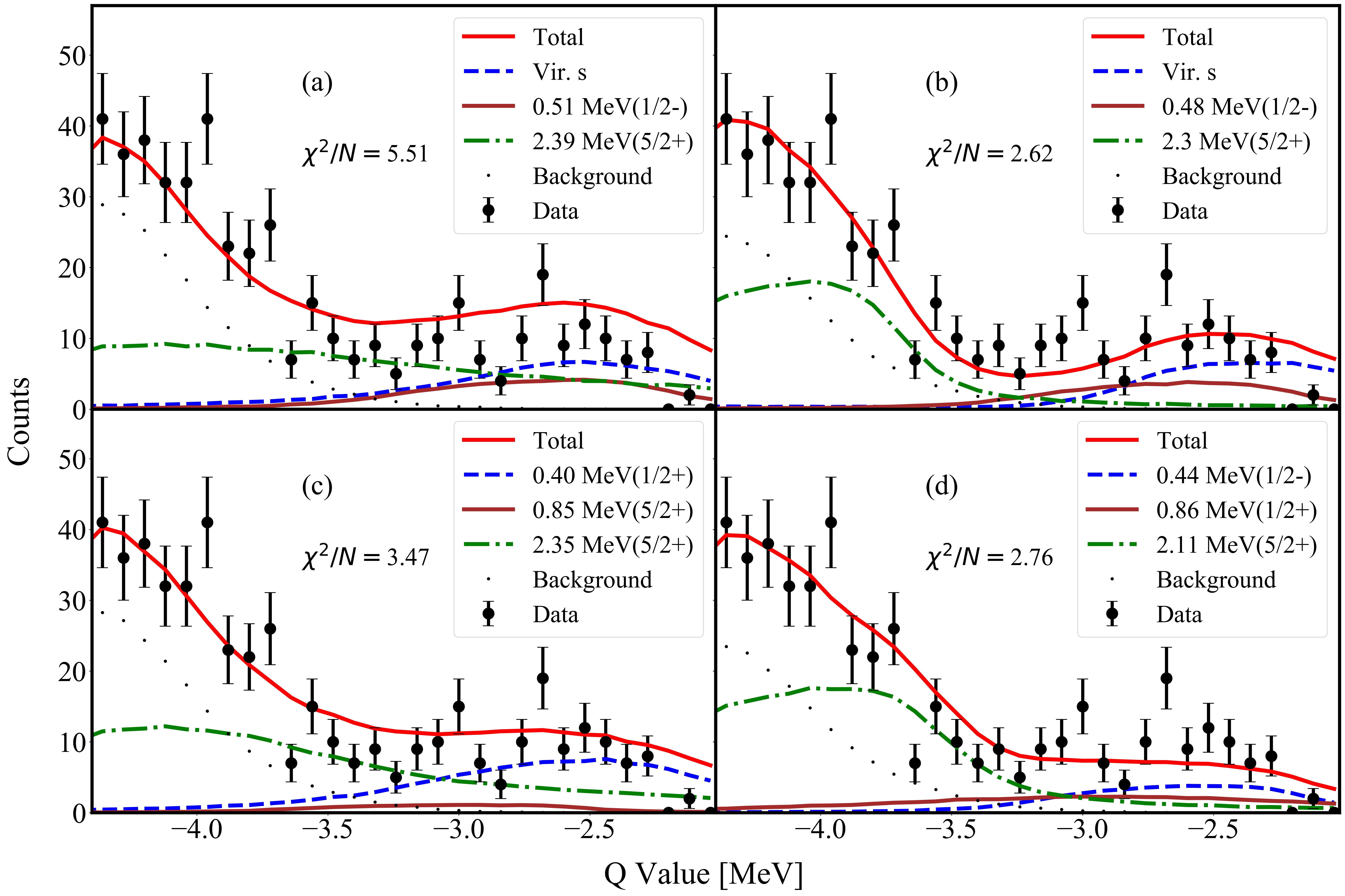}
\end{center}
\caption {Data fitted with GEANT4 simulations with energy and widths obtained from (a) \cite{Kon10}, (b) \cite{Cor19}, (c)  \cite{Ran14}, (d) \cite{Rib18}. The amplitudes of the states were used from the angular distributions. The global fit is shown as the red line and the background is denoted as black dots. The lowest-lying strength is shown as a blue-dashed line irrespective of its nature. The higher-lying states are depicted as solid brown and green dot-dashed lines. }\label{fig:2}
\end{figure}

To make a more robust comparison between the recent invariant-mass measurements and the transfer reaction, the resonance parameters extracted from \cite{Kon10},  \cite{Cor19},  \cite{Ran14}, and \cite{Rib18} were used as inputs for the GEANT4 \cite{Gea03} simulation, as shown in Figure \ref{fig:2}. The centroid energies and widths from the analyses were used and the relative intensities of the resonances and virtual states were fitted as free parameters. A relatively poor fit to the data from the $^{12}$Be(d,p) reaction experiment was produced from the simulations using resonance parameters from \cite{Kon10} ($\chi^2/\rm{NDF}=5.51$). Using the parameters from \cite{Ran14}  provided a better fit ($\chi^2/\rm{NDF}=3.47$) and those from \cite{Aks13} (not shown) gave a fit with $\chi^2/\rm{NDF}=3.18$. The parameters that gave the best fit from the literature were those from \cite{Cor19} ($\chi^2/\rm{NDF}=2.62$), closely followed by those from \cite{Rib18} ($\chi^2/\rm{NDF}=2.76$). These can all be compared to the Bayesian fit of the data with a single $p-$wave resonance along with the well-known $d$-wave resonance around 2~MeV. This fit, allowed the locations and widths of the two resonances to vary along with the intensities. The parameters noted in Table \ref{Exp} resulted in a $\chi^2/\rm{NDF}$ of $2.02$. The single $p$-wave below the 2~MeV $\frac{5}{2}^+$ resonance is the scenario that best agrees with these data. This dominance of $p$-wave strength near the particle threshold is in agreement with the results of \cite{Cas20}.

\section{Summary}
The beryllium chain of isotopes displays various clustering phenomena including molecular structures and one- and two-neutron halos in $^{11}$Be and $^{14}$Be respectively. The isotope $^{13}$Be is an unbound sub-system of the Borromean nucleus $^{14}$Be. Its structure has been investigated experimentally for forty years using both missing mass and invariant mass techniques. However, with the exception of the ~2~MeV $\frac{5}{2}^+$ resonance, the low-lying structure is still disputed. A new single-neutron transfer reaction experiment has brought new data and a new analysis technique involving GEANT4 simulations and a Bayesian fitting routine. The best fit of the Q-value data was obtained with a narrow, 0.55~MeV $p$-wave resonance and the $d$-wave resonance located at 2.22~MeV. Adding either a virtual state or a second resonance below 1~MeV produced somewhat poorer fits ($\chi^2/\rm{NDF}=3.0 - 3.4$) that were dominated by the $s$-wave contribution ($61\%-89\%$) with a small either $p$-($39\%$) or $d$-wave ($11\%$) resonance. Using the resonance parameters from either \cite{Cor19} or \cite{Rib18} produced better fits to the new data from the literature. The literature resonance parameters producing the poorer fits were those with a broader $\frac{5}{2}^+$ resonance \cite{Kon10, Ran14}.

\section*{Conflict of Interest Statement}
%All financial, commercial or other relationships that might be perceived by the academic community as representing a potential conflict of interest must be disclosed. If no such relationship exists, authors will be asked to confirm the following statement: 

The authors declare that the research was conducted in the absence of any commercial or financial relationships that could be construed as a potential conflict of interest.

\section*{Author Contributions}

KJ was the spokesperson for the new $^{12}$Be+$^2$H experiment and wrote the first draft of this manuscript. JK analyzed the data and produced the figures from the new $^{12}$Be+$^2$H experiment. RK was co-spokesperson for the new $^{12}$Be+$^2$H experiment and was instrumental in the design and running of the experiment. All authors contributed to the manuscript revision, read, and approved the submitted version.

\section*{Funding}
This research was supported by the U.S. Department of Energy, Office of Science, Office of Nuclear Physics under Contract No. DE-FG02-96ER40963 (UTK), DE-AC05-00OR22725 (ORNL), and the U. S. National Science Foundation under Award Numbers PHY-1404218 (Rutgers) and PHY-2011890 (Notre Dame). The authors are grateful for support from NSERC, Canada Foundation for Innovation and Nova Scotia Research and Innovation Trust, RCNP, the grant-in-aid program of the Japanese government. TRIUMF is supported by a contribution from the National Research Council, Canada. This work was supported by the National Research Foundation of Korea (NRF) grant funded by the Korean government (MSIT) Nos. 2020R1A2C1005981 and 2016R1A5A1013277. This work was partially supported by STFC Grant No. ST/L005743/1 (Surrey)

\section*{Acknowledgments}
The authors would like to acknowledge the collaborators on the TRIUMF $^{12}$Be+$^2$H experiment, and the beam delivery team for providing the $^{12}$Be beam.

%\section*{Supplemental Data}
 %\href{https://www.frontiersin.org/guidelines/author-guidelines#supplementary-material}{Supplementary Material} should be uploaded separately on submission, if there are Supplementary Figures, please include the caption in the same file as the figure. LaTeX Supplementary Material templates can be found in the Frontiers LaTeX folder.

\section*{Data Availability Statement}
The data from the recent $^{12}$Be+$^2$H measurement can be made available upon reasonable request.
% Please see the availability of data guidelines for more information, at https://www.frontiersin.org/guidelines/policies-and-publication-ethics#materials-and-data-policies

\bibliographystyle{Frontiers-Harvard} %  Many Frontiers journals use the Harvard referencing system (Author-date), to find the style and resources for the journal you are submitting to: https://zendesk.frontiersin.org/hc/en-us/articles/360017860337-Frontiers-Reference-Styles-by-Journal. For Humanities and Social Sciences articles please include page numbers in the in-text citations 
\bibliography{Jones_be_frontiers.bib}

%%% Make sure to upload the bib file along with the tex file and PDF
%%% Please see the test.bib file for some examples of references

%%% If you don't add the figures in the LaTeX files, please upload them when submitting the article.
%%% Frontiers will add the figures at the end of the provisional pdf automatically
%%% The use of LaTeX coding to draw Diagrams/Figures/Structures should be avoided. They should be external callouts including graphics.

\end{document}